\documentclass[]{spie}  

\usepackage{amsmath,amsfonts,amssymb}
\usepackage{graphicx}
\usepackage[colorlinks=true, allcolors=blue]{hyperref}
\usepackage{media9}

\title{Automated Kidney Segmentation by Mask R-CNN in T2-weighted Magnetic Resonance Imaging}

\author[a]{Manu Goyal}
\author[a]{Junyu Guo}
\author[a]{Lauren Hinojosa}
\author[a]{Keith Hulsey}
\author[a]{Ivan Pedrosa}
\affil[a]{Department of Radiology, 5323 Harry Hines Blvd., UT Southwestern Medical Center, Dallas, USA}

\authorinfo{Further author information: (Send correspondence to Manu Goyal)\\Manu Goyal.: E-mail: Manu.Goyal@UTSouthwestern.edu}

\begin{document} 
	\maketitle
	
	\begin{abstract}
		Despite the recent advances of deep learning algorithms in medical imaging, the automatic segmentation algorithms for kidneys in MRI exams are still scarce. Automated segmentation of kidneys in Magnetic Resonance Imaging (MRI) exams are important for enabling radiomics and machine learning analysis of renal disease. In this work, we propose to use the popular Mask R-CNN for the automatic segmentation of kidneys in coronal T2-weighted Fast Spin Eco slices of 100 MRI exams. We propose the morphological operations as post-processing to further improve the performance of Mask R-CNN for this task. With 5-fold cross-validation data, the proposed Mask R-CNN is trained and validated on 70 and 10 MRI exams and then evaluated on the remaining 20 exams in each fold. Our proposed method achieved a dice score of 0.904 and IoU of 0.822.  
	\end{abstract}
	
	
	\section{INTRODUCTION}
	\label{sec:intro}  
	
	Kidney cancer is among the 6th most common cancers in men and 8th in women, and the 5-year survival rate is 75\%\cite{acs, cdac}. However, there is wide variability in the prognosis among different renal cancers. Most kidney cancers (70\%) are first diagnosed as an incidental small renal mass (SRM; $\leq$4 cm) on an imaging study performed for an unrelated clinical reason. Furthermore, many of these renal masses are benign tumors that mimic cancers. Although radiomics may improve the characterization of renal masses, these analyses are often time-consuming. An automated deep learning system that provides automatic segmentation of kidney and renal masses would be a useful supporting tool for clinicians \citenum{dwivedi2021magnetic, souza2019automatic}. In recent times, artificial intelligence (AI) and deep learning methods such as convolution neural networks (CNN) have achieved good results in automated recognition of abnormalities in various medical imaging modalities such as X-ray, computed tomography (CT), positron emission tomography (PET), dermoscopy, ultrasound and Magnetic Resonance Imaging (MRI) \citenum{goyal2021sensitivity, hatt2018first, goyal2019skin, akkus2017deep, yap2020breast, ahmad2018semantic}. An AI segmentation tool for SRM is lacking and reports about automatic segmentation of the kidneys on MRI with AI algorithms have focused primarily on patients with adult polycystic kidney disease \citenum{kline2017performance}. Moreover, respiratory motion between different MRI acquisitions and between slices of the same acquisition makes AI more challenging.  
	
	We hypothesize that accurate segmentation of the normal renal parenchyma is the first step toward the development of fully automated AI algorithms for renal mass characterization. This work focuses on the validation of a Mask R-CNN for the automated segmentation of kidneys in T2-weighted (T2w) fast spin-echo (FSE) slices of MRI exams.

	\section{MRI Dataset}
	
	Our dataset consists of 100 MRI exams from patients with known renal masses imaged at different institutions and referred to our center for definitive treatment. We used the coronal 2D T2w FSE slices of each MRI exam as input for the proposed deep learning method. The number of slices in T2W MRI exams per patient varied from 11 to 56, and the total number of slices was 2423. 
	
	\subsection{Expert Annotations}
	\label{sec:title}
	
	All slices were annotated by a third-year radiology resident who created kidney masks per patient. All segmentation masks were reviewed by a senior radiologist for correction. These masks were used as ground truth in this study. There were a total of 1914 segmentation masks.  
	
	\section{Methodology}
	This section describes our proposed techniques for the detection of kidneys in T-2 weighted MRI exams. The preparation of dataset, deep learning approaches used for detection of kidneys in T-2 weighted MRI exams are detailed in this section. 
	
	\subsection{Pre-processing}
	We first pre-processed the MRI exams with the N4 bias correction algorithm. The height and width of MRI exams in this dataset varies between 256$\times$256 to 748$\times$748. We resized all the 2D slices to 512$\times$512.
	
	\subsection{Deep Learning Method}
\begin{figure}
		\centering
		\includegraphics[scale=.55]{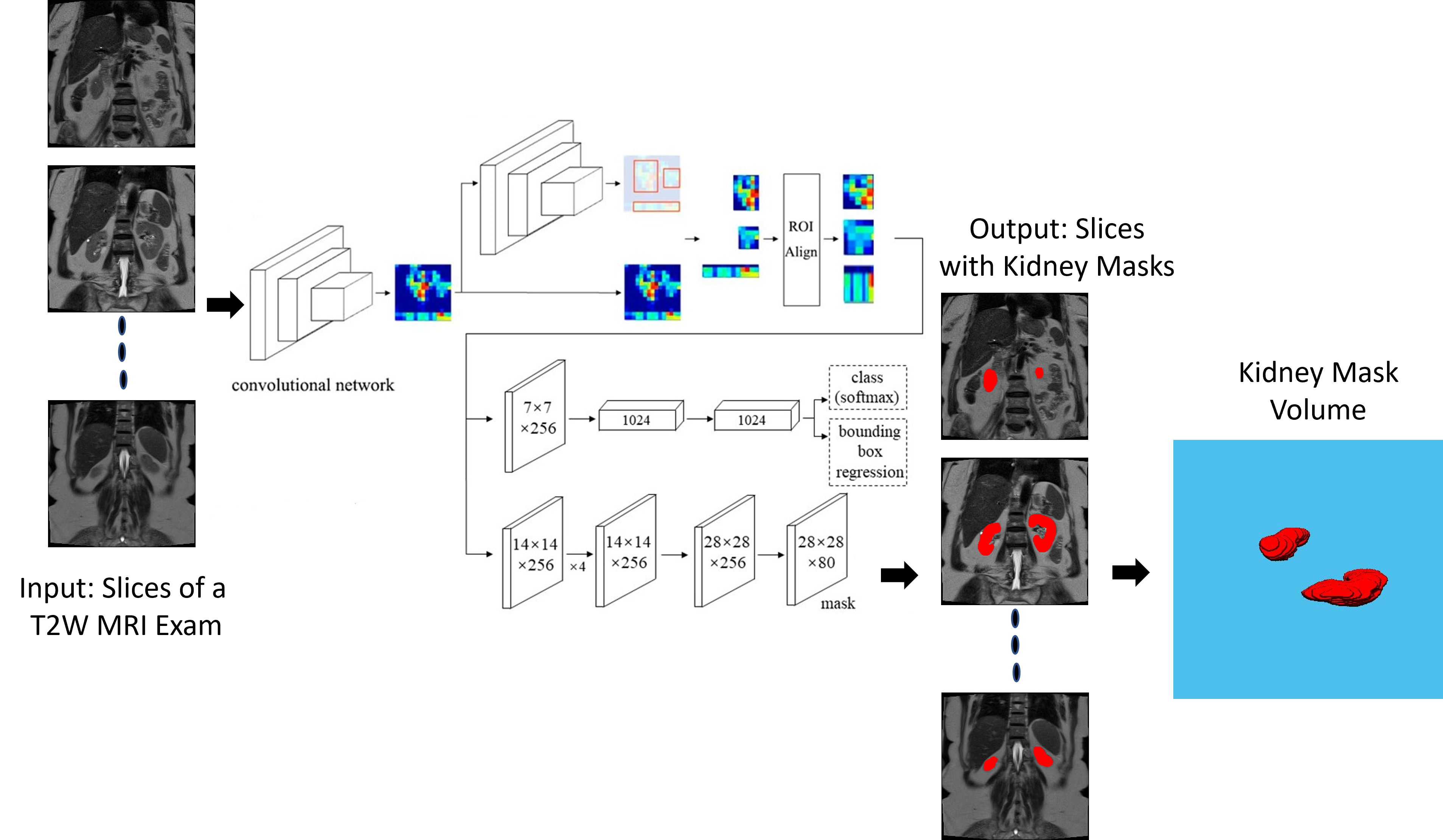}
		\caption{Architecture of Mask R-CNN for segmentation of kidney in T-2 weighted MRI exams}
		\label{fig:ROCM}
	\end{figure}

	Mask R-CNN is a deep learning architecture extended from Faster R-CNN, which determines the object's location by drawing a bounding box (detection) and then marking each pixel of the object with a mask (segmentation) \citenum{he2017mask}. In this work, we used the Mask R-CNN architecture with InceptionResNetV2 as the CNN network to segment kidneys in 2D T2W FSE slices and later combined the output masks to create kidney volume as shown in Figure \ref{fig:ROCM}.

	\subsection{Morphological Operations}
\begin{figure}
		\centering
		\includegraphics[scale=.55]{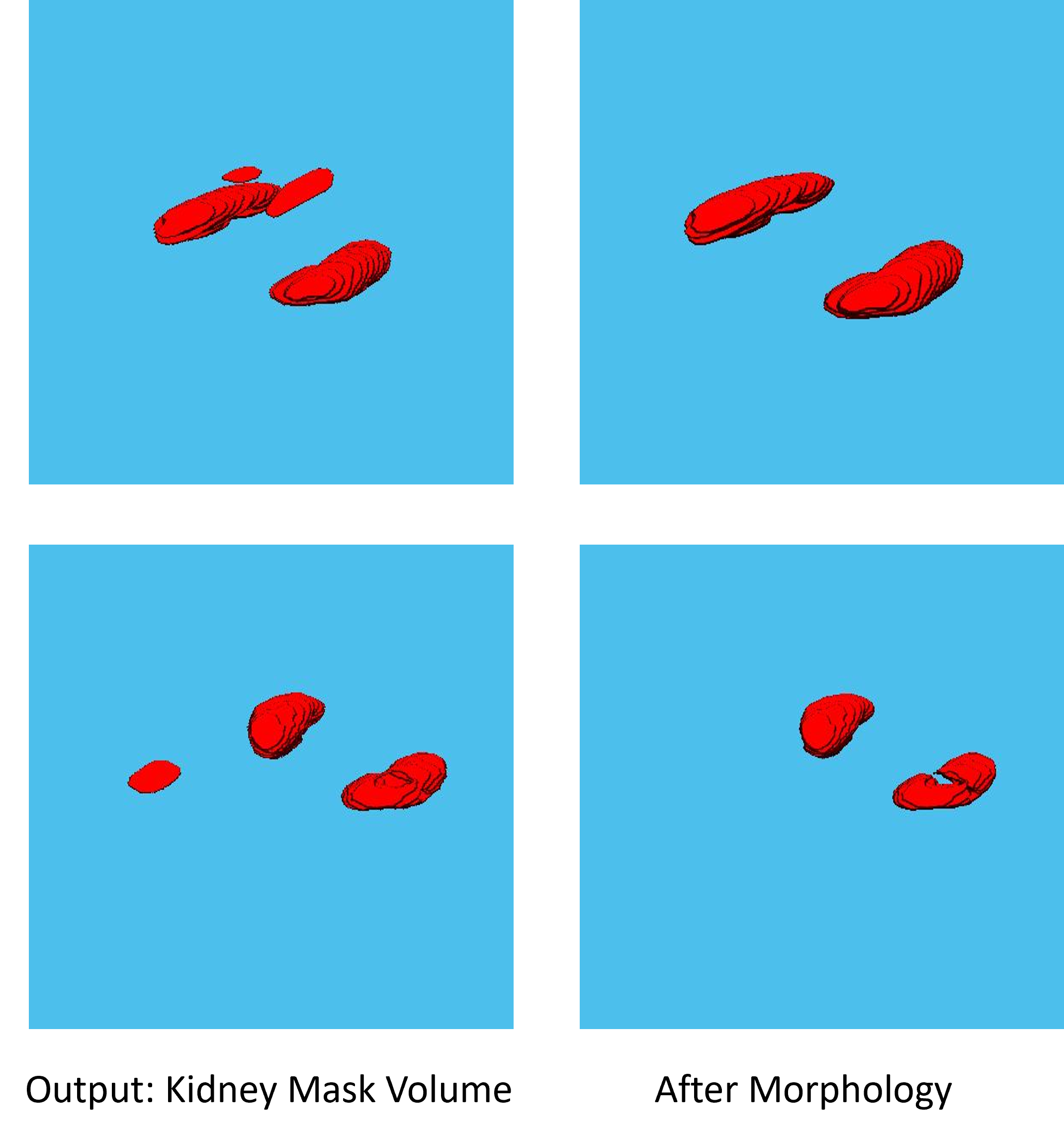}
		\caption{Comparison of Kidney Mask Volume before and after morphological operations}
		\label{fig:ROC}
	\end{figure}

	We used 3D morphological operations on binary volumes of kidney as the post-processing method to further refine the segmentation produced by the Mask R-CNN. We used the common dilation and erode morphological functions as well as clean operation (removing any random binary volume or voxel that is not connected to kidney), majority operation (retains the voxel if more than half of the voxels in the neighborhood) and fill operation (fill isolated interior voxels) as shown in the Figure \ref{fig:ROC}. 
	
	\subsection{Training, Validation and Test Set Splits}
	The Mask R-CNN method was trained on approximately 70 MRI exams as a training set and validated on 10 exams, and the remaining 20 exams were used as a testing set in a single fold. We evaluated the performance of Mask R-CNN on 5-fold cross-validation data to test the whole dataset.
	
	\subsection{Training Parameters}
	We used the TensorFlow object detection library to implement the Mask R-CNN InceptionResNetV2 method with 32 GB Nvidia Titan V100 GPU. We set a batch size of 4, the total number of epochs is 100, an initial learning rate of 0.008, with a momentum optimizer of 0.9, and gradient clipping by norm of 10. The final evaluation model is selected based on minimum validation loss. We also used the pre-trained model for fine-tuning and data augmentation techniques such as horizontal and vertical flips on the fly during training.

	\begin{table*}[]
		\centering
		\addtolength{\tabcolsep}{2pt}
		\renewcommand{\arraystretch}{1.5}
		\caption{Performance evaluation of our proposed methods for segmentation of kidneys in T-2 weighted MRI exams where MPP is Morphological Post-Processing}
		\label{resultst}
		\scalebox{0.75}
		{
			\begin{tabular}{lllllll} 
				\hline Method/ Evaluation Metrics  & Fold 1 & Fold 2 & Fold 3 & Fold 4 & Fold 5 & Average \\\hline \hline
				Mask R-CNN / Dice           &  0.890$\pm$0.038      & 0.904$\pm$0.022        &  0.871$\pm$0.057      &  0.891$\pm$0.048     & 0.892$\pm$0.042  & 0.890$\pm$0.044        \\
				Mask R-CNN/ IoU             &  0.804$\pm$0.059      &   0.825$\pm$0.036     & 0.776$\pm$0.089       & 0.807$\pm$0.074        & 0.837$\pm$0.062  & 0.810$\pm$0.069         \\
				Mask R-CNN with MPP/ Dice    &  0.902$\pm$0.037      & 0.914$\pm$0.019       &   0.886$\pm$0.058     & 0.910$\pm$0.041        & 0.910$\pm$0.038  &    0.905$\pm$0.041     \\
				Mask R-CNN with MPP/ Jaccard &  0.823$\pm$0.058      & 0.843$\pm$0.032        &  0.801$\pm$0.091      &    0.836$\pm$0.064    &  0.808$\pm$0.067      &    0.822$\pm$0.067     \\ \hline
		\end{tabular}}
	\end{table*}
	
		\begin{figure}
		\centering
		\includegraphics[scale=.95]{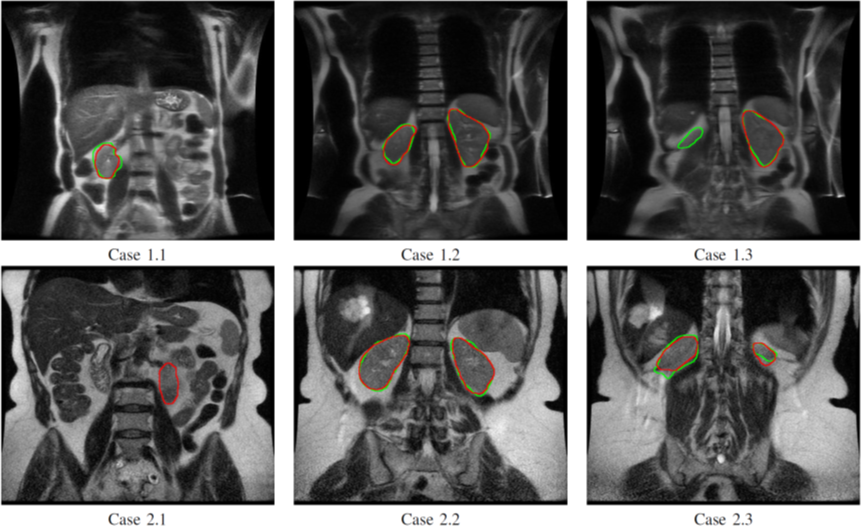}
		\caption{Comparison of ground truth (green) and inference (red) masks generated by Mask R-CNN with post-processing in 2D T2w FSE slices of two MRI exams. There is an FN in Case 1.3, an FP and a TP in Case 2.1, and all TP in other cases.}
		\label{fig:Results}
	\end{figure} 
	
	\begin{figure}
		\centering
		\includegraphics[scale=.75]{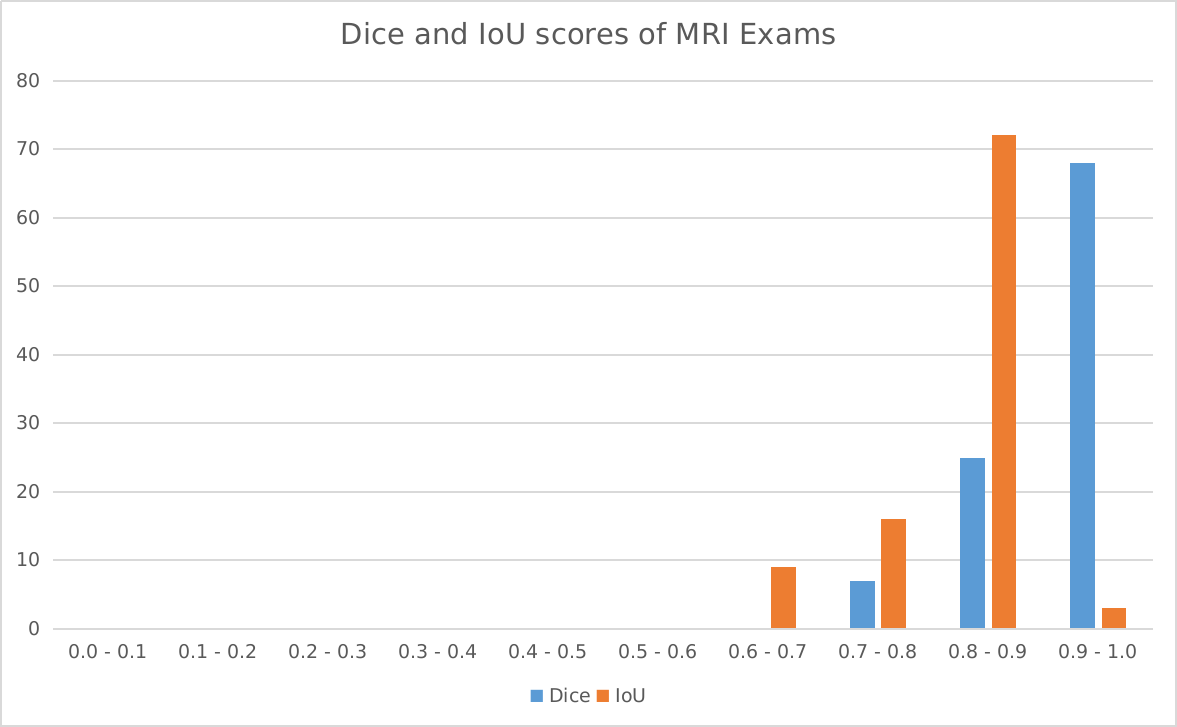}
		\caption{The bar plot demonstrates the total number of exams in which Dice and IoU scores ranged from 0 to 1.}
		\label{fig:DiceI}
	\end{figure}

	\section{Results}
	Our proposed method Mask R-CNN with morphological operations achieved the average dice score of 0.904 with a standard deviation (SD) of 0.041 and IoU or Jaccard Score of 0.822$$\pm$$0.066 across all folds, as shown in Table \ref{resultst}. When compared with the performance of original Mask R-CNN, our proposed post-processing methodology improved the segmentation performance by 1.5\% in dice score and 1.2\% in IoU.  The low SD suggests, the method performs consistently well across all the folds. The method detected the shape and location of the kidney correctly. Few examples of the correct and incorrect inferences are shown in Figure \ref{fig:Results}.

	The distribution of the number of cases in which dice and IoU scores ranged from 0 to 1 is shown in Figure \ref{fig:DiceI}. Out of 100 MRI exams used in this study, our proposed method achieved the dice score $\geq$ 0.8 in 93 cases as shown in Figure \ref{fig:DiceI}.

	\section{Conclusion}
	In this work, we validate the Mask R-CNN for the segmentation of kidneys in coronal T2W FSE MRI exams. Our method performed consistently well in all performance measures for 5-fold cross-validation data. The main limitation of this work is the small dataset. Further studies will be conducted to validate in larger datasets and other MRI sequences such as T1-weighted.
	
	Our proposed method accurately segmented the kidney, hence, provides us an automated application to assess the location and size of the kidney in MRI exams. It is also an essential step in applying textural analysis radiomics where the kidney segmentation is required to aid in the diagnosis of abnormalities \citenum{de2019radiomics, krajewski2018imaging}.

	\bibliography{report} 
	\bibliographystyle{spiebib} 
	
\end{document}